\documentclass[12pt,preprint]{aastex}
\usepackage{epsfig}

\newcommand{\cir}{CIR X-1}

\slugcomment{Version of 11 March 2003}

\shorttitle{Long-term X-ray variability of \cir}
\shortauthors{Saz Parkinson et al.}

\begin{document}

\title{Long-term X-ray variability of Circinus X-1}

\author{P. M. Saz Parkinson\altaffilmark{1}, 
D. M. Tournear, E. D. Bloom, W. B. Focke, K. T. Reilly} 
\affil{Stanford Linear Accelerator Center, Stanford University, Stanford, CA 94309}
\vspace{0.7mm}
\vspace{0.5mm}
\author{K. S. Wood, P. S. Ray, M. T. Wolff}
\vspace{0.5mm}
\affil{E. O. Hulburt Center for Space Research, 
    Naval Research Laboratory,
    Washington, DC 20375}
\vspace{0.7mm}
\and
\author{Jeffrey D. Scargle}
\affil{NASA/Ames Research Center, Moffett Field, CA 94035}
\altaffiltext{1}{pablos@SLAC.stanford.edu}


\begin{abstract}
We present an analysis of long term X-ray monitoring observations of Circinus X-1 (Cir X-1) made with four different instruments: Vela 5B, Ariel V ASM, Ginga ASM, and RXTE ASM, over the course of more than 30 years. We use Lomb-Scargle periodograms to search for the $\sim$16.5 day orbital period of Cir X-1 in each of 
these data sets and from this derive a new orbital ephemeris based solely on X-ray measurements, which we compare to the previous ephemerides obtained from radio observations. We also use the Phase Dispersion Minimization (PDM) technique, as well as FFT analysis, to verify the periods obtained from periodograms. We obtain dynamic periodograms (both Lomb-Scargle and PDM)
of Cir X-1 during the RXTE era, showing the period evolution of Cir X-1, and also displaying some unexplained discrete jumps in the location of the peak power. 
\end{abstract} 

\section{Introduction}

Cir X-1 was discovered in 1971 \citep{1971ApJ...169L..23M} though it was most likely detected as early as 1965 in a survey conducted by Friedman and collaborators at NRL \citep{1967Sci...156..374F}. Observations made with the Uhuru satellite from 1971 to 1973 \citep{1974ApJ...191L..71J} suggested that it was a binary source with a period longer than 15 days. Observations by Ariel V over 1975-76 resulted in the determination of a 16.6 day period \citep{1976ApJ...208L..71K}. A binary model was proposed in which the high state occurs near the periastron of a highly elliptical orbit \citep{1975Natur.254..674C}. Cir X-1 was initially classified as a BHC
due to its temporal and spectral similarities to Cyg X-1: millisecond
variability \citep{1977ApJ...215L..57T}, flickering in the hard state
and very soft energy spectrum in the high state. It has also been found to display a high energy ($>$ 20keV) power law tail which is considered a signature of BHCs \citep{2001ApJ...547..412I}. The observation by EXOSAT of Type I X-ray bursts \citep{1986MNRAS.221P..27T} in 1985, however, led 
to the classification of Cir X-1 as a neutron star. It is normally classified as a Low Mass X-ray Binary (LMXB), although doubts still remain about the spectral type of the companion \citep{1992A&A...260L...7M}, with some authors proposing an intermediate, 3-5 $M_{\odot}$ companion \citep{2001MNRAS.328.1193J}. The orbital period of Cir X-1 is longer than any other LMXB \citep{1995xrb..book.....L}.
The recent discovery of P-Cygni profiles \citep{2000ApJ...544L.123B} implies the presence of strong winds, which those authors interpret to be coming from an X-ray heated accretion disk, implying that Cir X-1 is probably being viewed edge-on. Radio observations show that Cir X-1 displays relativistic jet-like emission \citep{1998ApJ...506L.121F} leading to its classification as a microquasar. These jet-like features appear to trail back towards SNR G321.9-0.3 \citep{1998ApJ...506L.121F,1999MNRAS.310.1165T} causing some to speculate that Cir X-1 is the remains of a young (age $\sim$20,000--100,000 years), asymmetric supernova. Recent HST observations \citep{mignani}, however, rule out any association between Cir X-1 and SNR G321.9-0.3, thereby eliminating the age constraint imposed by such an association and raising the possibility that Cir X-1 could be a much older system.

In this paper, we look at archival observations of Cir X-1 made by Vela 5B, Ariel V ASM, Ginga ASM, and RXTE ASM. The data span over 30 years (from 1969 May to 2002 September) and therefore allow us to test for small changes in the orbital period. We first use periodograms to search for the orbital period in the four different sets of data. From these observations we derive a new orbital ephemeris of Cir X-1, based solely on these X-ray observations (all previous ephemerides were based on radio
observations). We then look at dynamic periodograms of the more than 6.5 years of RXTE ASM data. We show that there is a clear coherent periodicity in the flux at the orbital period. Furthermore, we show that this modulation disappears for an extended period of time, during which a separate modulation at $\sim$40 days is detected. This new peak in the periodogram lasts for almost one year, after which the fundamental 16.5 day peak returns (and the 40 day peak disappears).

\section{Observations}
We used a series of observations taken by four different instruments over the course of the last $\sim$33 years. We chose those instruments which had an ASM and therefore allowed us to study the long-term behavior of the source. Figure \ref{fig1} shows the composite light curve obtained by incorporating all the observations by the different instruments into one single plot. All the data have been normalized to the Crab. Included in this plot are the EXOSAT observations which displayed the Type I X-ray bursts to illustrate what the flux level of the source was at the time, though in this paper we do not analyze the EXOSAT data.
Figure \ref{fig2} shows the light curves obtained by the four ASM instruments, individually plotted for greater clarity. In the short paragraphs that follow I briefly describe the four instruments for which data was analyzed in this paper.

\subsection{Vela 5B}

The Vela 5B satellite \citep{1969ApJ...157L.157C} was placed in orbit on 1969 May 23 and operated until 1979 June 19, although telemetry tracking was poor after mid-1976. The X-ray detector covered the celestial sphere twice per orbit and had an effective area of 26 cm$^2$, with an energy range of 3-12 keV. In our analysis we use 485 observations of Cir X-1 made between MJD 40368 and MJD 42692.

\subsection{Ariel V}

Ariel V was launched on 1974 October 15 and operated until 1980 March 14. The All Sky Monitor (ASM) experiment \citep{1976Ap&SS..42..123H} consisted of two small X-ray pinhole cameras with an effective area of $\sim$1 cm$^2$ and an energy range of 3-6 keV. In this paper we look at 1173 observations of Cir X-1 made between MJD 42345 and MJD 42692.

\subsection{Ginga ASM}

The Ginga observatory was launched on 1987 February 5 and reentered the Earth's atmosphere on 1991 November 1. The ASM instrument \citep{1989PASJ...41..391T} consisted of two proportional 
counters operating in the 1--20 keV energy range, covered by six different collimators, each with a field of view of roughly 1$^{\circ}\times$45$^{\circ}$ (FWHM) and an effective area of $\sim$70 cm$^2$ (for a total effective area of $\sim$420 cm$^2$). For more details of the Ginga ASM see \cite{1989PASJ...41..391T}. In this paper, we analyze 300 observations of Cir X-1 taken between the dates of MJD 46855 and MJD 48532.

\subsection{RXTE ASM}

The Rossi X-ray Timing Explorer (RXTE) \citep{1993A&AS...97..355B}, was launched on 1995 December 30. The ASM instrument \citep{1996ApJ...469L..33L} consists of 3 Scanning Shadow Cameras (SSCs) with a total effective area of 90 cm$^2$ and a sensitivity to 1.5--10 keV X-rays in three energy channels (roughly corresponding to 1.5--3, 3--5, and 5--12 keV). Both 
individual ``dwell'' by ``dwell'' (where a ``dwell'' refers to a 90 s observation) and one-day average measurements are made available by the ASM/RXTE team via their website. The data used in this paper spanned the dates MJD 50088 (6 January 1996) through MJD 52536 (19 September 2002) and consisted of 1972 daily-averaged measurements.

\section{Data Analysis and Results}

We computed the Lomb-Scargle periodograms \citep{1982ApJ...263..835S} of the four different data sets. These are shown in Figure \ref{fig3}. A fiducial period of 16.6 days is shown on each plot with dashed lines, for comparison between the different epochs. We found peaks in our periodograms for Vela 5B data (16.69 $\pm$ 0.015 days), Ariel V (16.66 $\pm$ 0.01 days), Ginga ASM (16.577 $\pm$ 0.001 days), and RXTE ASM (16.5427 $\pm$ 0.0001 days). We computed the uncertainty on the period through Monte Carlo simulations of 1000 light curves, assuming the errors on the counting rate have a gaussian distribution. After computing the periodograms of the 1000 light curves generated, we took the mean and standard deviation of the peak frequency. 

We also used the Phase Dispersion Minimization (PDM) technique \citep{1978ApJ...224..953S} and compared the values obtained from this method with the periods obtained from the Lomb-Scargle periodograms. Our results are shown in Figure \ref{fig4}: Vela 5B data (16.67 $\pm$ 0.01 days), Ariel V (16.644 $\pm$ 0.001 days), Ginga ASM (16.565 $\pm$ 0.004 days), and RXTE ASM (16.5421 $\pm$ 0.0001 days).

Finally, Fourier analysis was performed on each of the archival data sets to search for orbital periodicities. A detailed review of Fourier analysis techniques in X-ray timing is given by \cite{Fourier}. We used a bin time of 7 days for the Fast Fourier Transform (FFT). This gives a Nyquist frequency of $8.267 \times 10^{-7}$, or a minimum period of 14.0 days. The analysis was performed by oversampling the FFT. This process creates an FFT with finer detail than a non-oversampled FFT. However, not all of the bins are independent. This method allows one to estimate the frequency of the maximum power in a power density spectrum (PDS) to better accuracy. In \cite{1976PhDT.........9M} it is shown that if the peak power in the signal occurs in frequency bin $k$ (where $k$ need not be an integer), then the best estimate of the signal frequency is given by

\begin{equation}
\hat{\nu}_{0} = \nu_{k} + \frac{3}{4\pi^{2}\epsilon
T}\left(\frac{\bar{P}_{k+\epsilon}-\bar{P}_{k-\epsilon}}{\bar{P}_k}\right)
\end{equation}
where $\bar{P}$ is normalized power, $T$ is the length of the time series, and
$n=1/\epsilon$ is the oversampling factor, for this analysis we used $n=8$.  The uncertainty on the frequency determination \citep{1976PhDT.........9M} is given by
\begin{equation}
\sigma_{\hat{\nu}_0}\approx\frac{1}{2\pi T}\sqrt{\frac{6}{P_k}}
\end{equation}

The Vela, Ariel and Ginga data sets were each FFT'd as a single data set. The RXTE ASM data set, however, was split into two separate data sets. The first data set spans the dates MJD 50088 through 50829, while the second data set covers the dates from MJD 51575 to MJD 52536. The reason for splitting up the observations is that there is a span of almost two years during which the orbital modulation is not clearly detectable. Starting at around MJD 50830, the period (and power) of the peak intensity in the FFT starts to rapidly decrease until it becomes undetectable. While it reappears soon after MJD 51200, there are other unexplained phenomena in the spectra which prevent a good measurement of the period. We describe this in more detail in section 3.2 of the paper. It is only after MJD 51575 that the orbital modulation once again becomes unambiguously detectable. Table \ref{tab:periodbyfft} shows the details and results of the FFT analysis, which are also plotted in Figure \ref{fig5}).

\begin{table}
\begin{center}
\begin{tabular}{|l|c|c|c|c|} \hline
\textbf{Experiment}     & \textbf{Start MJD}    & \textbf{Stop MJD}     &
\textbf{Period (days)} & \textbf{error}\\   \hline\hline
Vela    & 40367 & 42687 & 16.677 &$\pm 0.028$\\ \hline
Ariel   & 42345 & 44306 & 16.664 &$\pm 0.029$\\ \hline
Ginga   & 46855 & 48531 & 16.551 &$\pm 0.028$\\ \hline
RXTE ASM set 1 & 50088 & 50829 & 16.528 &$\pm 0.063$\\ \hline
RXTE ASM set 2 & 51575 & 52536 & 16.502 &$\pm 0.019$\\  \hline
\end{tabular}
\vspace*{+0.5cm}
\caption{Period searching results from FFT analysis.} \label{tab:periodbyfft}
\end{center} \end{table}
We also used epoch folding techniques to determine the period, but our results were inconclusive. This was most likely due to the changing shape of the pulse profile (see for example Figure \ref{fig7} for the change in pulse profile in the RXTE era).

\subsection{Orbital Ephemerides}
Several ephemerides have been published for Cir X-1 over the years. Figure \ref{fig5} shows some of those most often cited in the literature. \cite{nicolson} published an ephemeris based on radio flares that took place between 1976-1980. This ephemeris gave the following equation for the onset of the Nth onset of flares (phase 0):

$\mathrm{MJD_{N}=43,076.26 + N (16.588 - 1.66\times 10^{-4} N)}$, giving a \.{P}$=-2.00 \times 10^{-5}$ (labelled Nicolson 1980 in Figure \ref{fig5}). In the same IAUC, Nicolson reports that using the mean X-ray/radio period of 16.594 days derived for 1976-1977 he obtains the following ephemeris:

$\mathrm{MJD_{N}=43,076.26 + N (16.594 - 2.5\times 10^{-4} N)}$, giving a \.{P}$=-3.01 \times 10^{-5}$ (labelled Nicolson 1980 b in Figure \ref{fig5})

A more recent ephemeris, also calculated by Nicolson, though reported by \cite{1991MNRAS.253..212S}, was based on the onset times of radio flares observed between 1978 and 1988. This gives the following time for phase 0:

$\mathrm{MJD_{N}=43,076.37 + N (16.5768 - 3.53\times 10^{-5} N)}$, which yields a \.{P}$=-4.26 \times 10^{-6}$ (labelled Stewart 1991 in Figure \ref{fig5}). 

Figure \ref{fig5} also shows the different data points obtained by applying three different techniques to the four separate data sets. The individual Lomb-Scargle periodograms for each instrument are shown Figure \ref{fig3}. The PDM periodograms for each data set are shown in Figure \ref{fig4}. We also include in Figure \ref{fig5} the points obtained by using an over sampled FFT (labelled 'Period by FFT'). 

We use all the measurements obtained from the various data sets and period-finding techniques to determine a best fit. We fit the period to the function:

$\mathrm{P=P_{0}+}$\.{P}$\mathrm{(T-T_{0})}$ where we use $\mathrm{T_{0}=43,076.37}$. 

Our best fit parameters are: 

$\mathrm{P_{0}=16.6534\pm5.6\times10^{-3}}$ and \.{P}$\mathrm{=-1.6261\times10^{-5}\pm1.109\times10^{-6}}$. The function of best fit is also plotted in Figure \ref{fig5}. This \.{P} gives us a characteristic time scale of P/2\.{P} $\sim$1400 years, which is around a factor of $\sim$4 shorter than that obtained from the previous ephemeris (Stewart 1991).

We suppose that the ephemeris is quadratic (i.e. constant \.{P}) and therefore follows the equation (for small \.{P}):

$\mathrm{MJD_{N}=MJD_{0}+P_{0}N+\frac{1}{2}P_{0}}$\.{P}$\mathrm{N^2}$ where N is the cycle number and $\mathrm{MJD_{N}}$ is the Nth occurrence of phase 0. We use the same $\mathrm{MJD_{0}}$ as the previous ephemeris, which combined with the values of $\mathrm{P_{0}}$ and \.{P} obtained from our fit give us our new ephemeris:

$\mathrm{MJD_{0}=43,076.37 + N (16.6534 - 1.354\times10^{-4} N)}$

\subsection{Dynamic periodograms of the RXTE ASM data}

Using the RXTE ASM daily averaged data, we performed ``dynamic'' periodograms by selecting segments of data of 
length $\sim$200 days and then stepping through the entire data set in increments of 20 days. Each periodogram was 
taken in the frequency range between 0.02 and 0.2 d$^{-1}$ (5 to 50 day period). Although we are oversampling 
our data in this way and are therefore not producing statistically independent periodograms, this technique 
nevertheless allows us to observe long-term trends in the time-frequency space which would otherwise not be so 
clearly visible. We tried several different values for the length of our periodogram, as well as for the step. 
While these parameters changed the resolution of the figure, the overall features and their time scales remained the same.  

The left side of Figure \ref{fig6} shows a dynamic Lomb-Scargle periodogram, while the plot on the right in the same 
figure shows a dynamic PDM periodogram. Both plots share many of the same features: the $\sim$16.5 day orbital 
period is clearly visible in both, along with some harmonics (at 1/2 and 1/3 the orbital period). The PDM dynamic 
periodogram also shows several sub-harmonics (at 2 and 3 times the orbital period) which are picked out naturally by  
this period searching technique \citep{1978ApJ...224..953S}.
A key feature of both plots is the suppression of the 16.5 day peak starting at $\sim$MJD 50830 and lasting 
for around 270 days, until $\sim$MJD 51100. During this time, the most significant peak in both periodograms is 
centered at $\sim$40 days. 

To make sure that this feature is not an artifact of our periodogram technique, we look at its manifestation in the 
time domain. We split up the RXTE ASM data into ten separate segments, each roughly of $\sim$250 days in length. 
Figure \ref{fig7} shows the different segments of data have folded at the ``orbital'' period of 16.54 days. The 
profiles show that in the first several segments of data, the light curve is clearly modulated at the orbital 
period. However, when we get to the fourth segment (MJD 50839--51081), there is no sign of the modulation. 
Furthermore, if we take the fourth segment of data and fold it at 40 days, instead of the 16.54, then we see a 
clear modulation. Figure \ref{fig8} shows this segment of data folded at 40 days. The inset plot shows the 
Lomb-Scargle periodogram for the same data, with a clear peak at 40 days (and no detectable peak at $\sim$16.5 days). 
To further check these results, we took the 10 individual light curves shown in Figure \ref{fig7} and folded 
them at a period of 40 days. As expected, only the fourth segment shows a significant modulation at this period, 
and none of the other nine show this modulation.

The dynamic periodogram also shows an important peak at around 22 days. This peak is present for $\sim$200 
days before the 16.5 day peak is suppressed and right before the 40 day peak appears. When the 16.5 day 
orbital signal finally returns, it starts off at a period slightly shorter than 16.5 days and takes almost 
100 days to ramp back up to its previous 16.5 day peak. Finally, there is another feature common to both dynamic 
periodograms. At around MJD 51410, the 16.5 signal splits into two separate signals, one slightly longer period 
($\sim$18.4 days) and one slightly shorter ($\sim$15.9 days). This split lasts for $\sim$165 days, until both 
signals finally converge back into one at around MJD 51575. This last feature roughly coincides with the point 
in the light curve where the overall flux of the source begins its steady decline (see Figure \ref{fig2}). 
After several years in a high flux state, where its baseline flux was greater than 1 Crab, Cir X-1 has begun 
a steep decline in flux, towards a lower intensity state; a decline which is still in progress.

\section{Discussion}
\begin{figure}
\epsscale{1.0}
\plotone{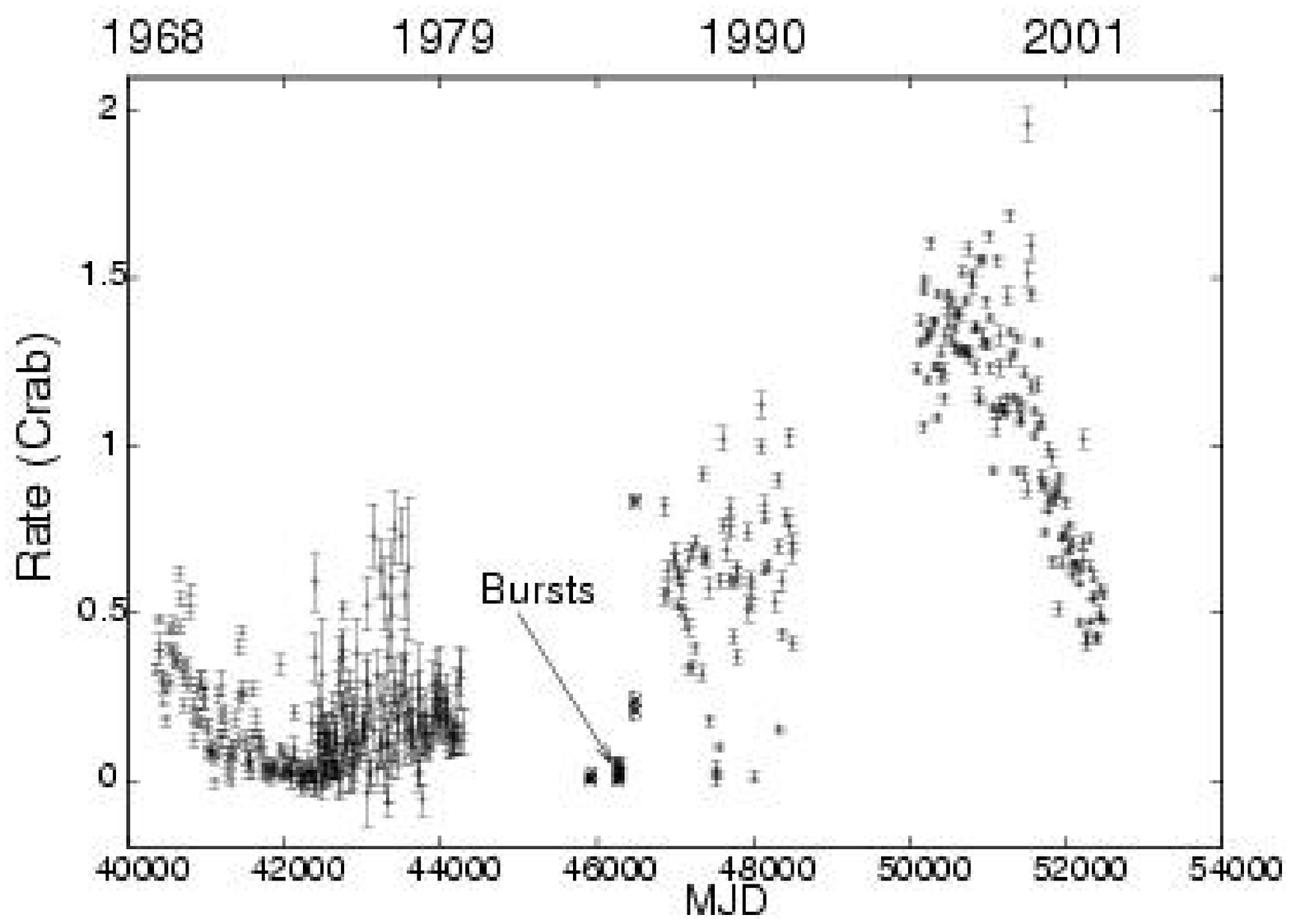}
\caption{Long-term Light Curve of Cir X-1. Fluxes have been normalized to the Crab. The figure includes data (in chronological order) from Vela 5B, Ariel V ASM, EXOSAT, Ginga ASM, and RXTE ASM. The arrow points to the EXOSAT observations which displayed Type I X-ray bursts.\label{fig1}}  
\end{figure}

\begin{figure}
  \begin{center}
    \begin{tabular}{cc}
	\resizebox{75mm}{!}{\rotatebox{270}{\plotone{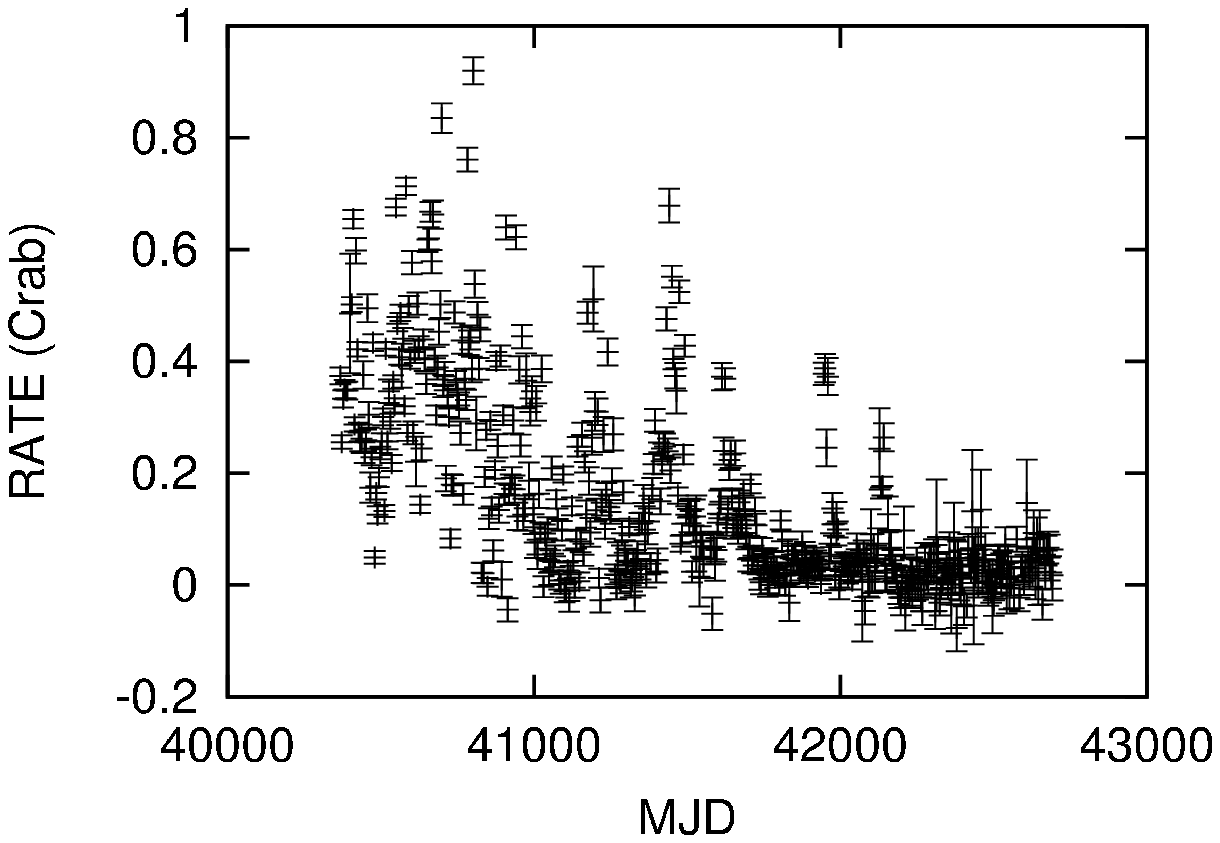}}} &
      	\resizebox{75mm}{!}{\rotatebox{270}{\plotone{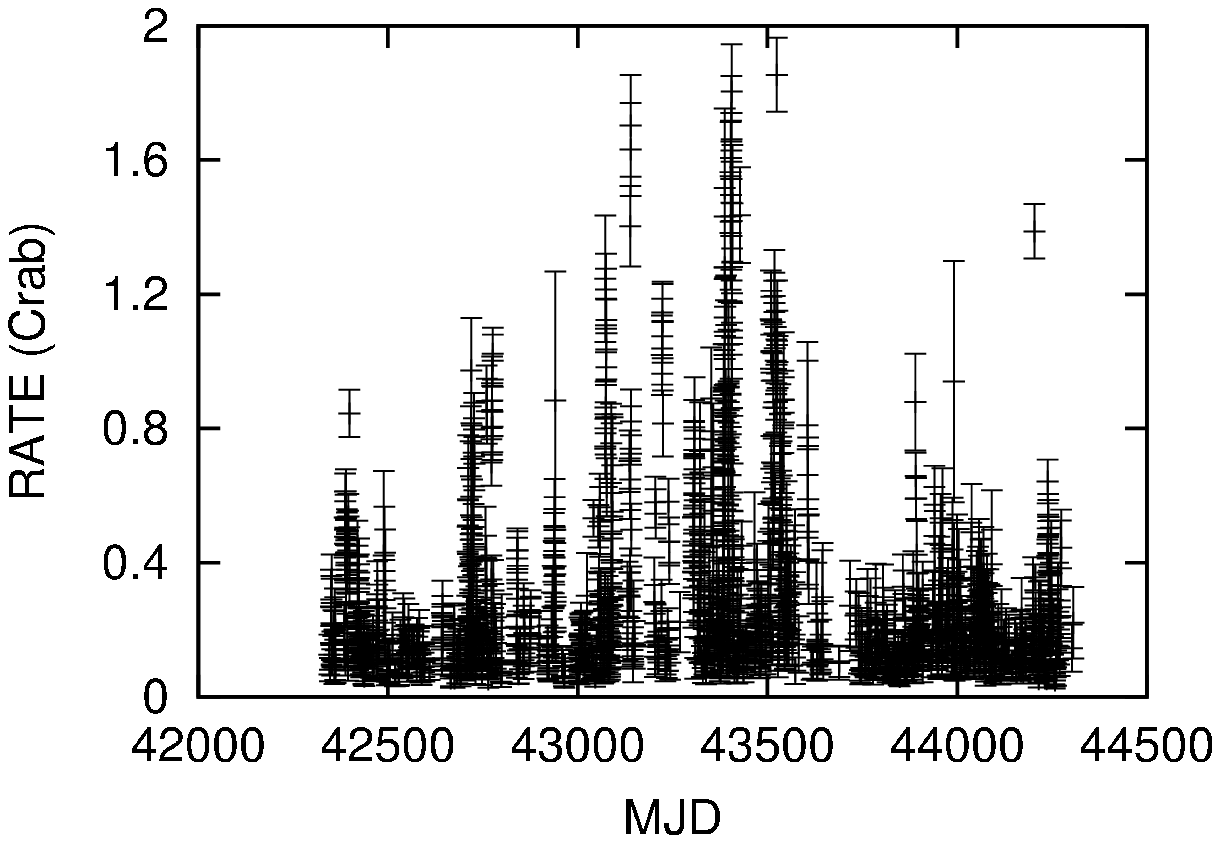}}} \\
      	\resizebox{75mm}{!}{\rotatebox{270}{\plotone{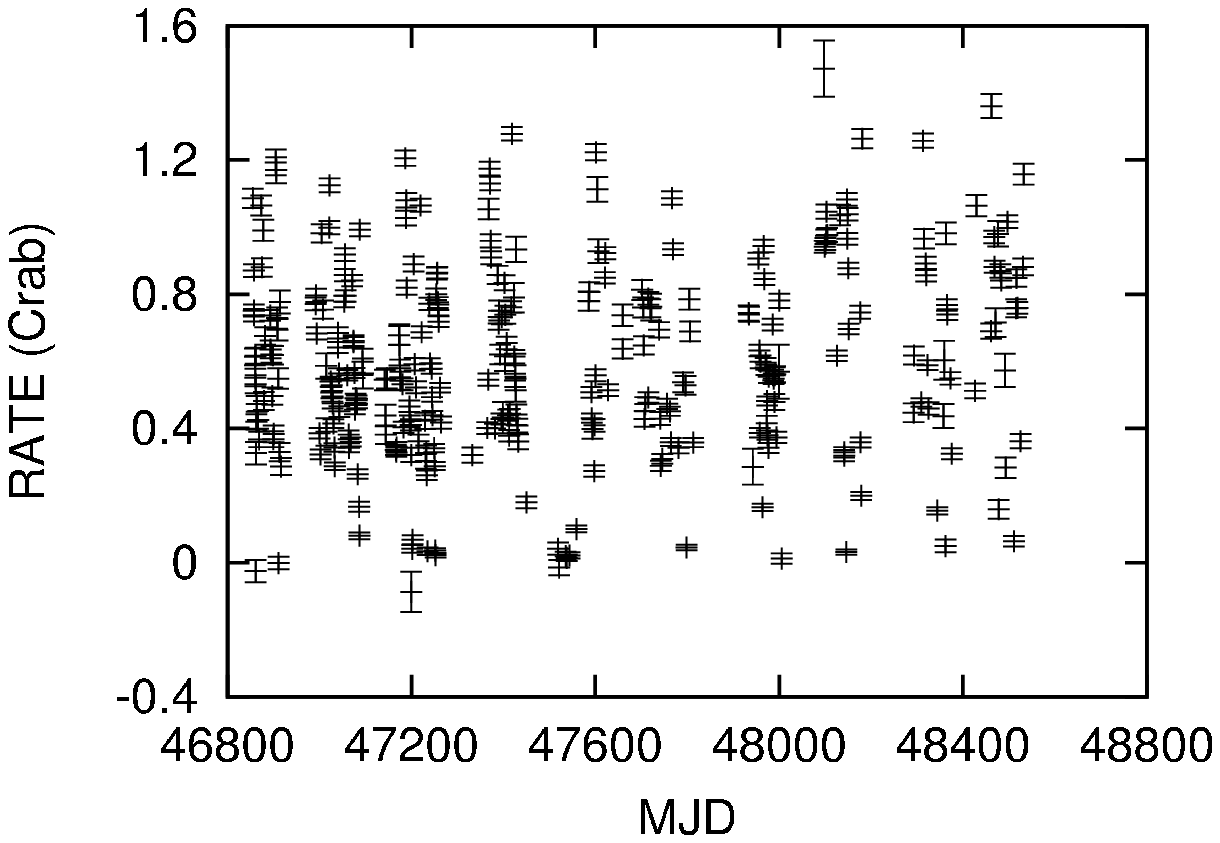}}} &
      	\resizebox{75mm}{!}{\rotatebox{270}{\plotone{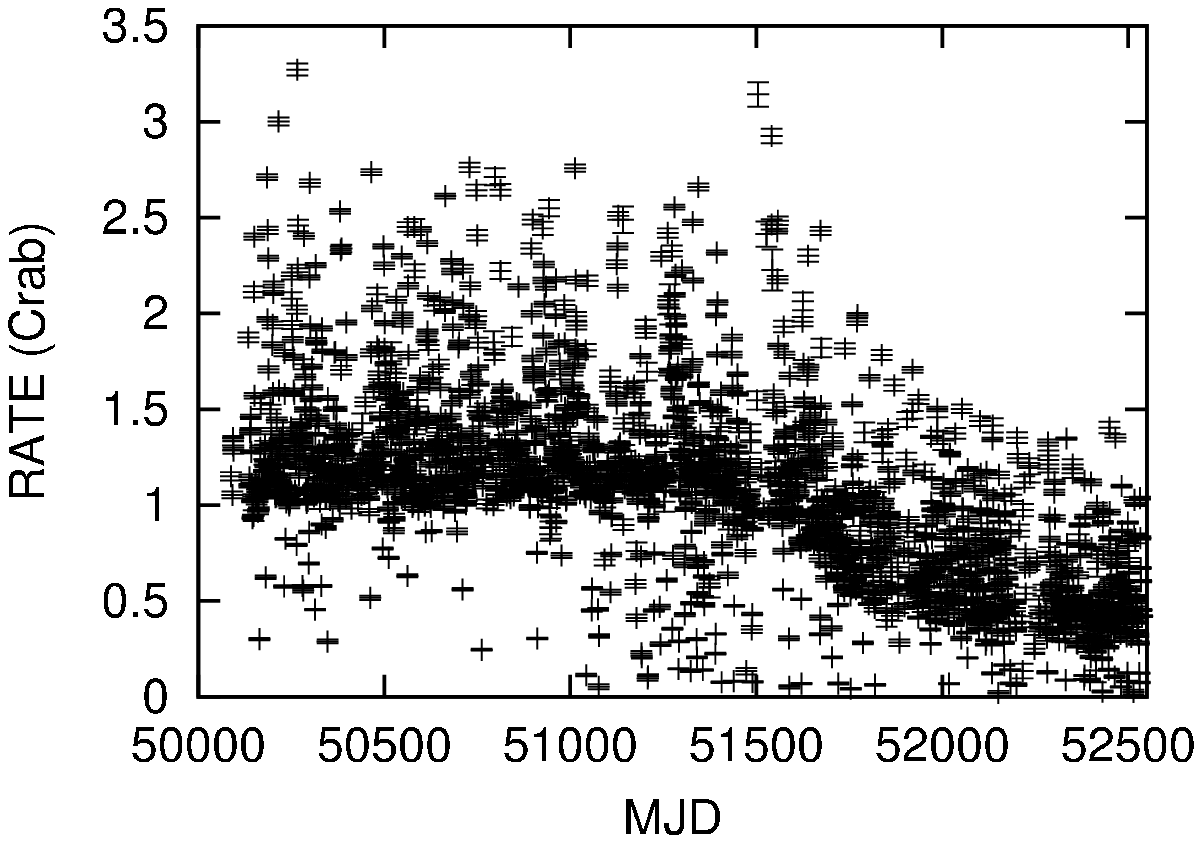}}} \\
    \end{tabular}
\caption{Individual light curves of Cir X-1 obtained by various instruments. Fluxes have been normalized
to the Crab. {\bf Top Left} -- Vela 5B (26 May 1969 - 7 October 1975). {\bf Top Right} -- Ariel V ASM (25 October 1974 - 8 March 1980). {\bf Bottom Left} -- Ginga ASM (1 March 1987 - 2 October 1991). {\bf Bottom Right} -- RXTE ASM (6 January 1996 - 19 September 2002).\label{fig2}}  
\end{center}
\end{figure}

\begin{figure}
  \begin{center}
    \begin{tabular}{cc}
	\resizebox{75mm}{!}{\rotatebox{270}{\plotone{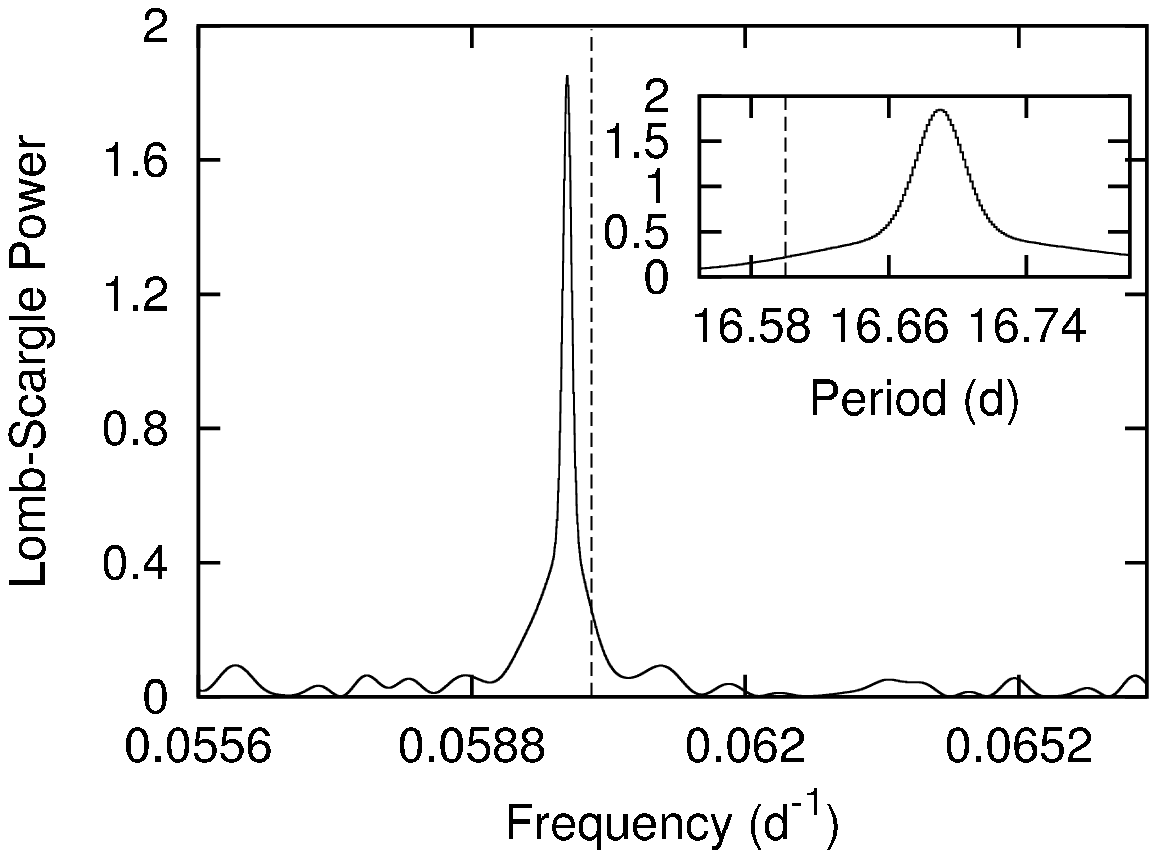}}} &
      	\resizebox{75mm}{!}{\rotatebox{270}{\plotone{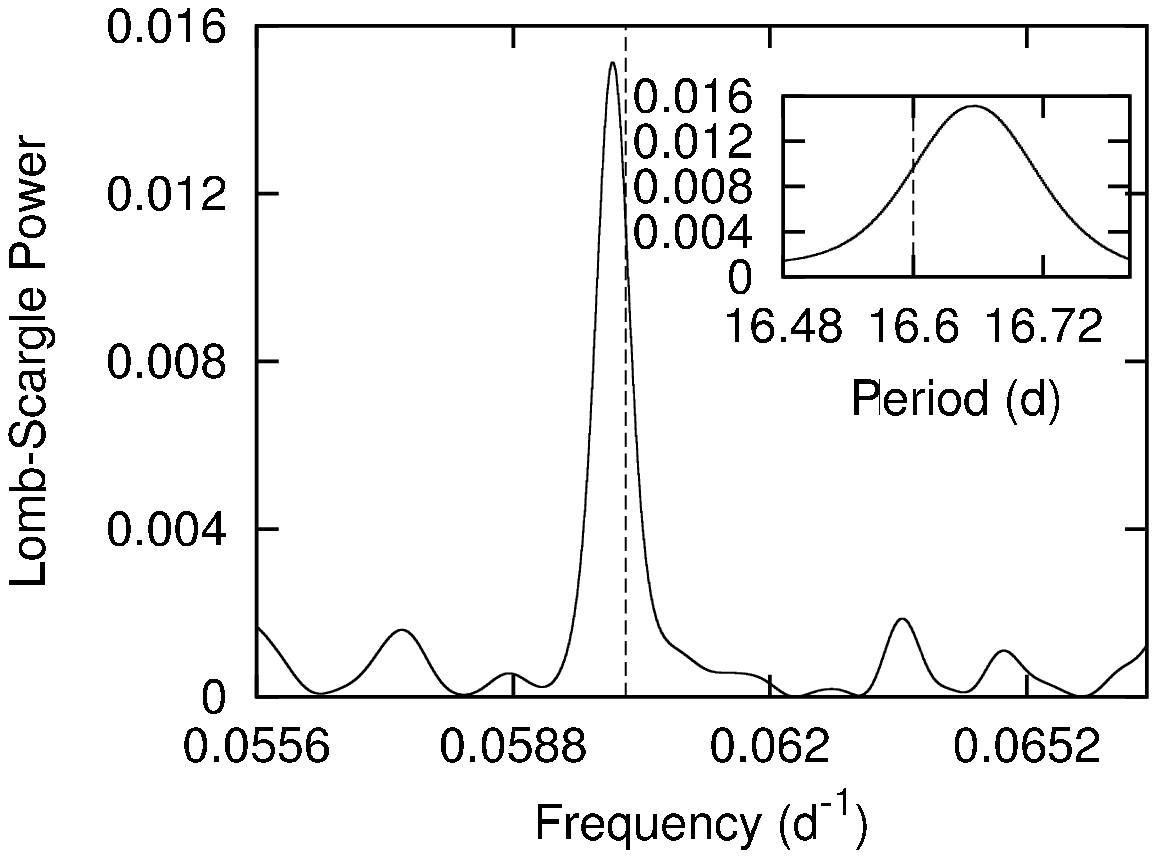}}} \\
      	\resizebox{75mm}{!}{\rotatebox{270}{\plotone{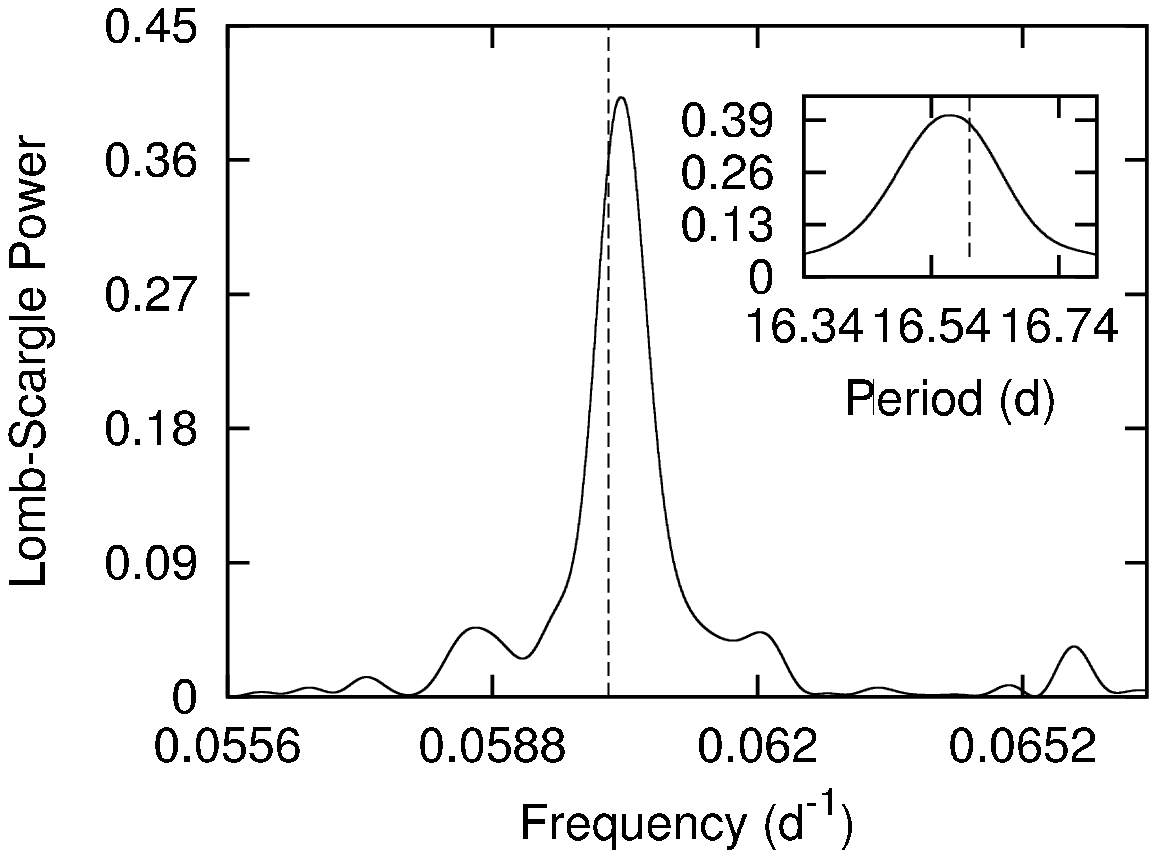}}} &
      	\resizebox{75mm}{!}{\rotatebox{270}{\plotone{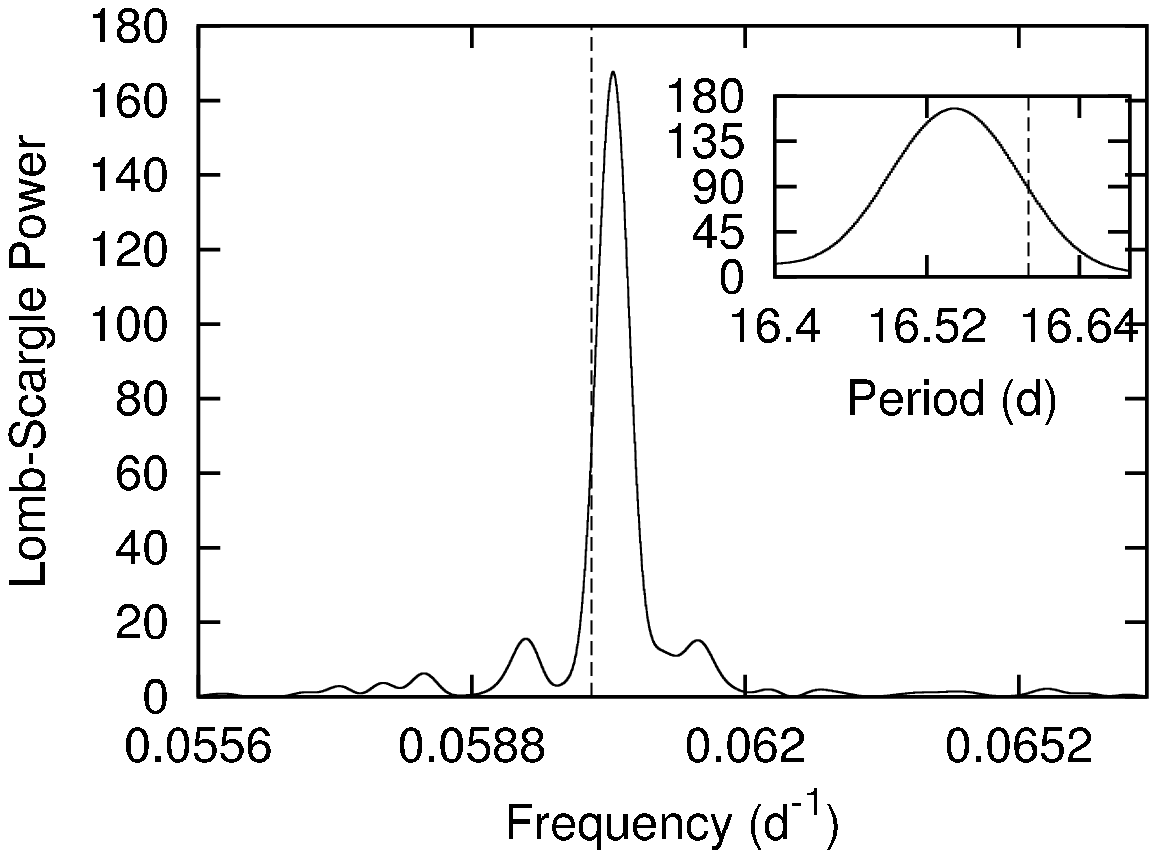}}} \\
    \end{tabular}
\caption{Lomb-Scargle periodograms of Cir X-1. The periodograms correspond to the data shown in the light curves in Figure \ref{fig2}. Inset plots show region around the orbital period in greater detail (with the x axis in units of time, for greater clarity). A fiducial period of 16.6 days is shown (with dashed lines) on each plot for comparison between epochs. {\bf Top Left} -- Vela 5B: Peak at 16.69 days. {\bf Top Right} -- Ariel V ASM: Peak at 16.66 days. {\bf Bottom Left} -- Ginga ASM: Peak at 16.57 days. {\bf Bottom Right} -- RXTE ASM: Peak at 16.54 days.\label{fig3}}  
\end{center}
\end{figure}

\begin{figure}
  \begin{center}
    \begin{tabular}{cc}
	\resizebox{75mm}{!}{\rotatebox{270}{\plotone{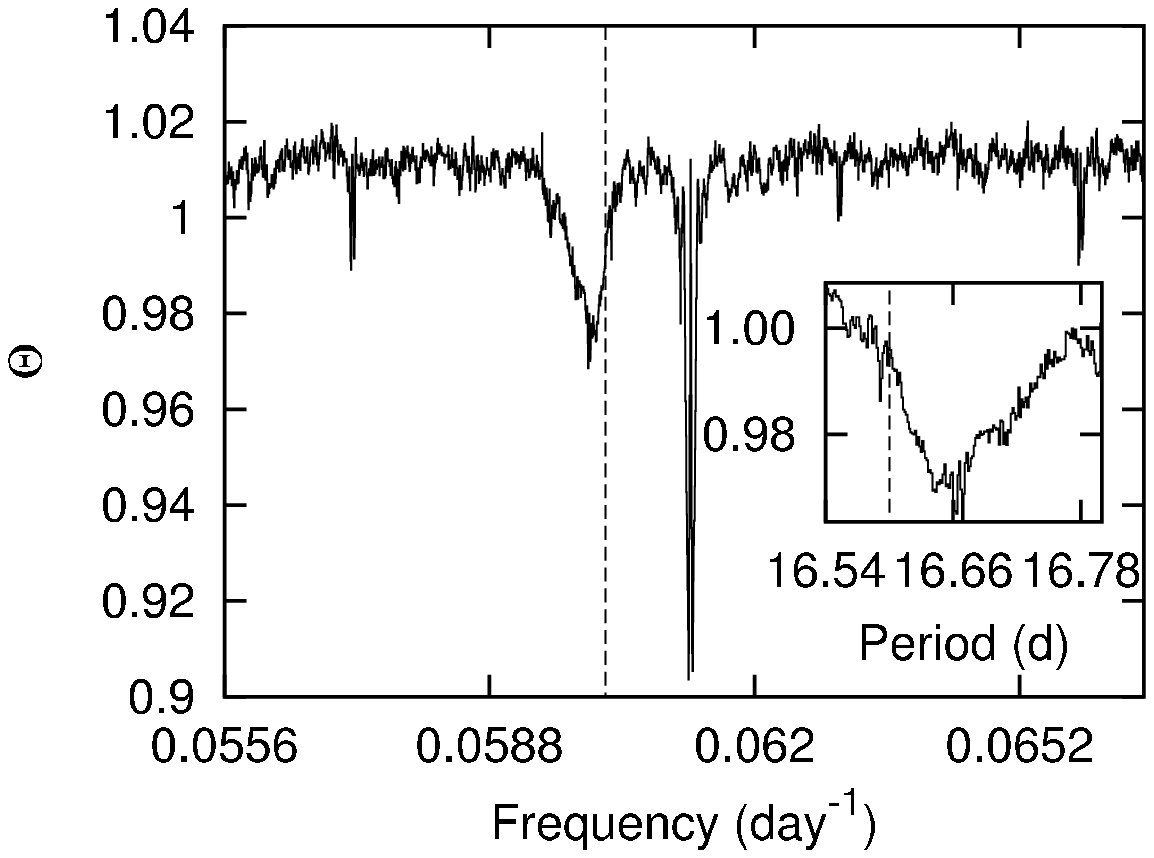}}} &
      	\resizebox{75mm}{!}{\rotatebox{270}{\plotone{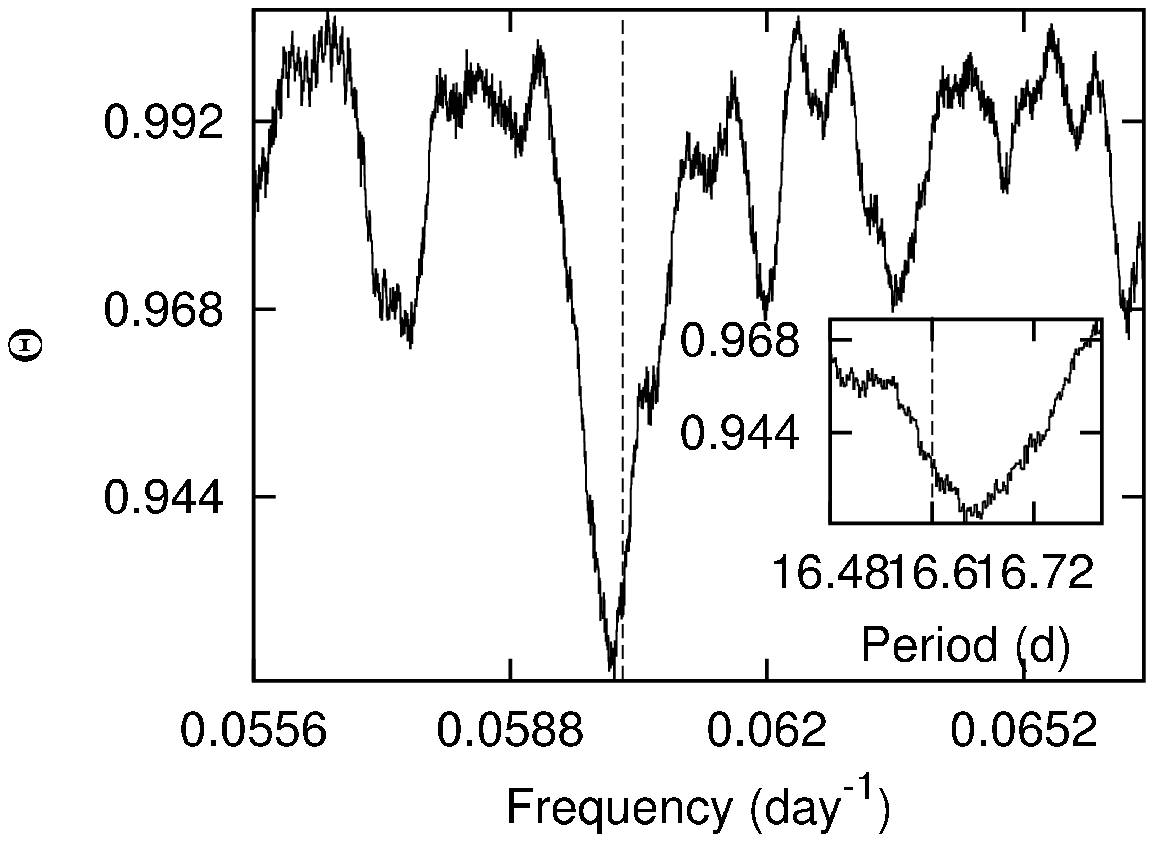}}} \\
      	\resizebox{75mm}{!}{\rotatebox{270}{\plotone{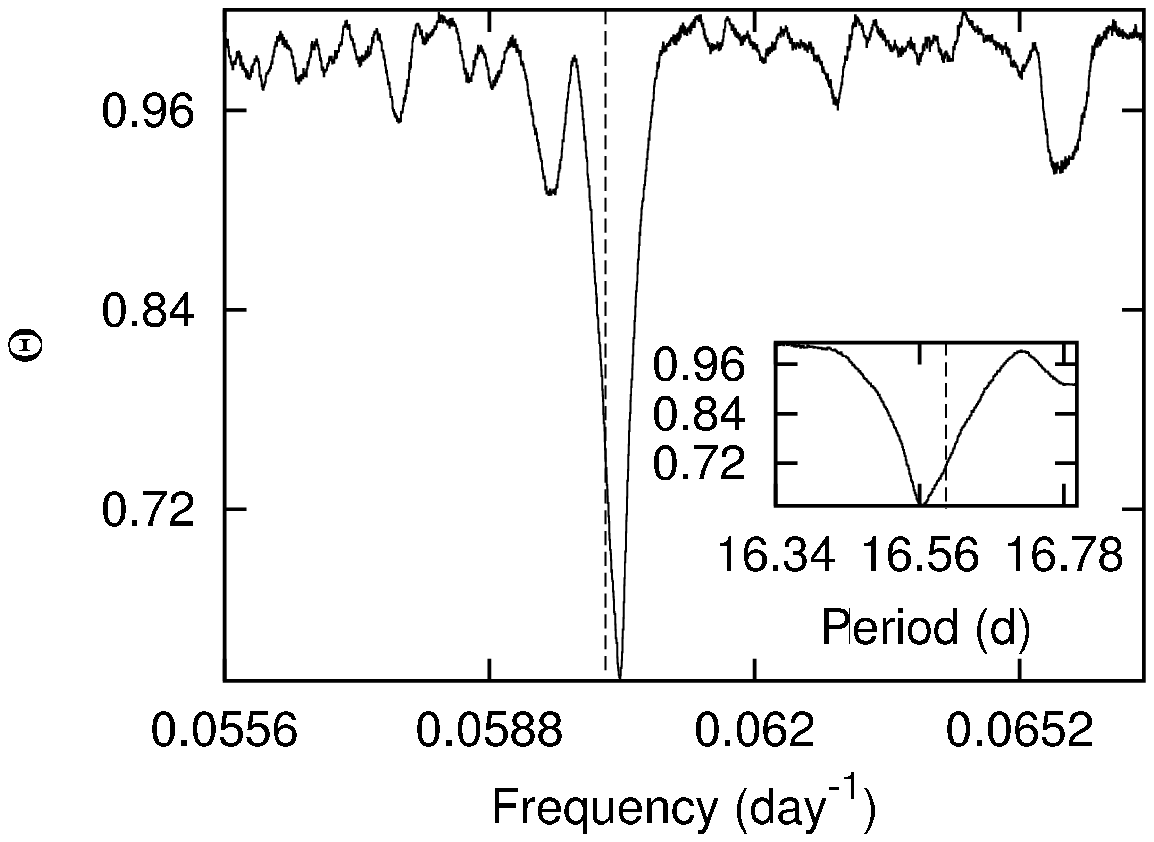}}} &
      	\resizebox{75mm}{!}{\rotatebox{270}{\plotone{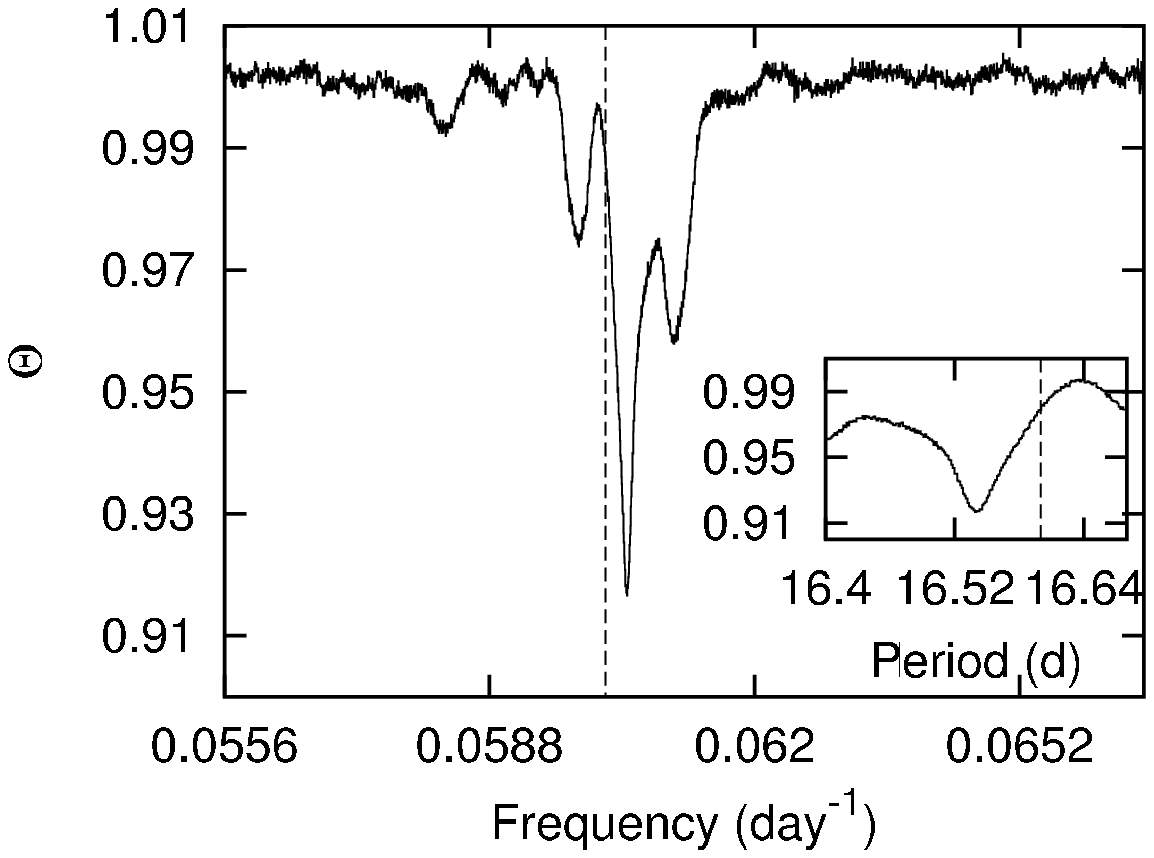}}} \\
    \end{tabular}
\caption{PDM periodograms of Cir X-1. The periodograms correspond to the data shown in the light curves in Figure \ref{fig2}. Inset plots show region around the orbital period in greater detail (with the x axis is in units of time, for greater clarity). A fiducial period of 16.6 days is shown (with dashed lines) on each plot for comparison between epochs. {\bf Top Left} -- Vela 5B: $\Theta_{min}$ at 16.33 days, but a second prominent period is present at $\Theta$=16.67 days. {\bf Top Right} -- Ariel V ASM: $\Theta_{min}$ at 16.64 days. {\bf Bottom Left} -- Ginga ASM: $\Theta_{min}$ at 16.56 days. {\bf Bottom Right} -- RXTE ASM: $\Theta_{min}$ at 16.54 days.\label{fig4}}
\end{center}
\end{figure}

\begin{figure}
\epsscale{1.0}
\plotone{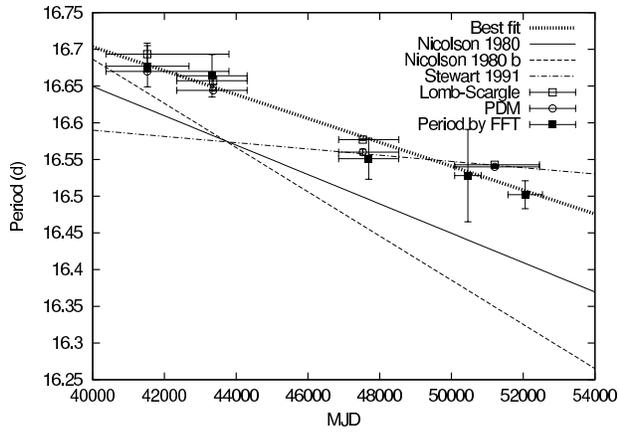}
\caption{Evolution of the orbital period of Cir X-1. The data points represent the best estimate for the orbital period based on Lomb-Scargle periodograms of the Vela 5B, Ariel V ASM, Ginga ASM, and RXTE ASM data, as well as the PDM periodograms, and the FFT technique (period by FFT).\label{fig5}}  \end{figure}

Cir X-1 has been extensively monitored now for over 30 years, though there have been long gaps in the observations, 
as can be seen from Figure \ref{fig1}. The first thing to note about the long-term light curve shown in 
Figure \ref{fig1} is the large variation in the source intensity. Ignoring the large variability within the 16.6 
day orbit (itself very significant), we see that the overall flux level of the source has gone from about 0.4 Crab 
at the beginning of the Vela 5B era, down to almost zero a few years later, then increasing at every epoch since 
then, until reaching a peak of more than 1 Crab in the RXTE era, only to begin its decline once again at $\sim$MJD 51600, 
a decline which appears to have levelled off at about 0.4 Crab. 
Included in Figure \ref{fig1}, besides the data from the four observatories which had an ASM, are the individual 
pointed observations from EXOSAT. It is interesting to note that the Type I X-ray bursts which led to the 
classification of Cir X-1 as a neutron star \citep{1986MNRAS.221P..27T} were all detected during one EXOSAT 
observation taken on 1985 August 12 ($\sim$MJD 46290) at a time when Cir X-1 was much dimmer than it has been 
ever since. The mean rate at the time of the first Type I burst was $\sim$15 cts$\,$s$^{-1}$ and at the time of 
the other two it was $\sim$90 cts$\,$s$^{-1}$ (see \cite{1986MNRAS.221P..27T} for more details); this corresponds 
to $\sim$1--5\% of the Crab. This has been offered as an explanation as to why no more bursts have ever been observed. 
However, it appears that Cir X-1 could soon have a flux similar to its 1984-85 level, making it an interesting 
potential target for detection of new Type I X-ray bursts.

In this paper we have attempted to derive a long-term ephemeris for Cir X-1 based solely on X-ray observations. 
Using the results we obtained from Lomb-Scargle periodograms, PDM periodograms, and FFT analyses, we derived a 
line of best fit which provides, as far as we know, the first ephemeris solely derived from X-ray observations. 
Although the X-ray emission of Cir X-1 is correlated with the radio observations \citep{1991MNRAS.253..212S}, the 
peak in X-ray emission sometimes precedes and sometimes follows the onset of the radio flares \citep{1998PhDT........31S}. 
It would be interesting to compare our newly derived X-ray ephemeris with a more recent one obtained from radio 
observations, unfortunately since the late 1980's the radio flares from Cir X-1 have been too weak to update the 
ephemeris \citep{1998PhDT........31S}.

\begin{figure}
\epsscale{1.0}
\plottwo{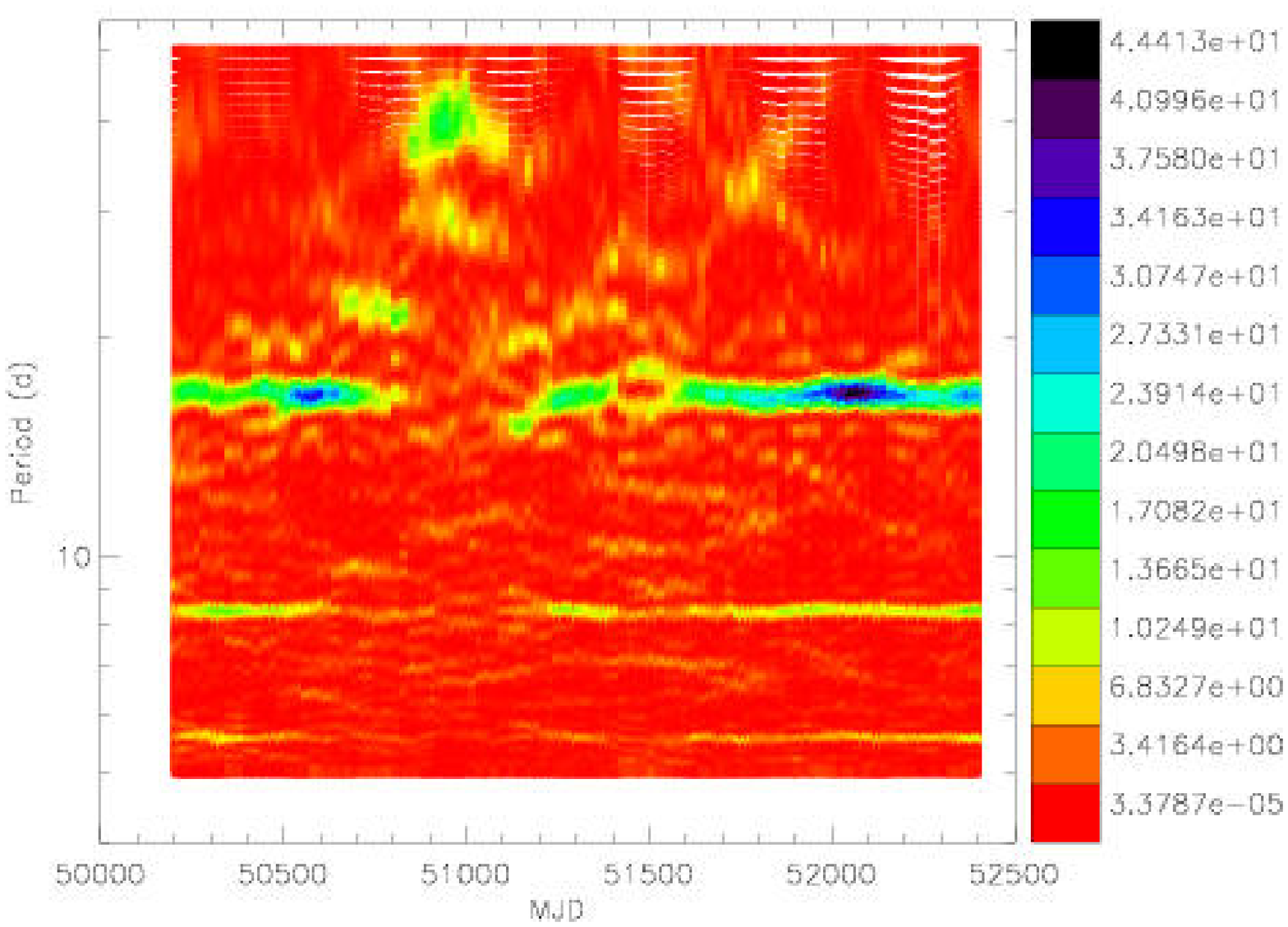}{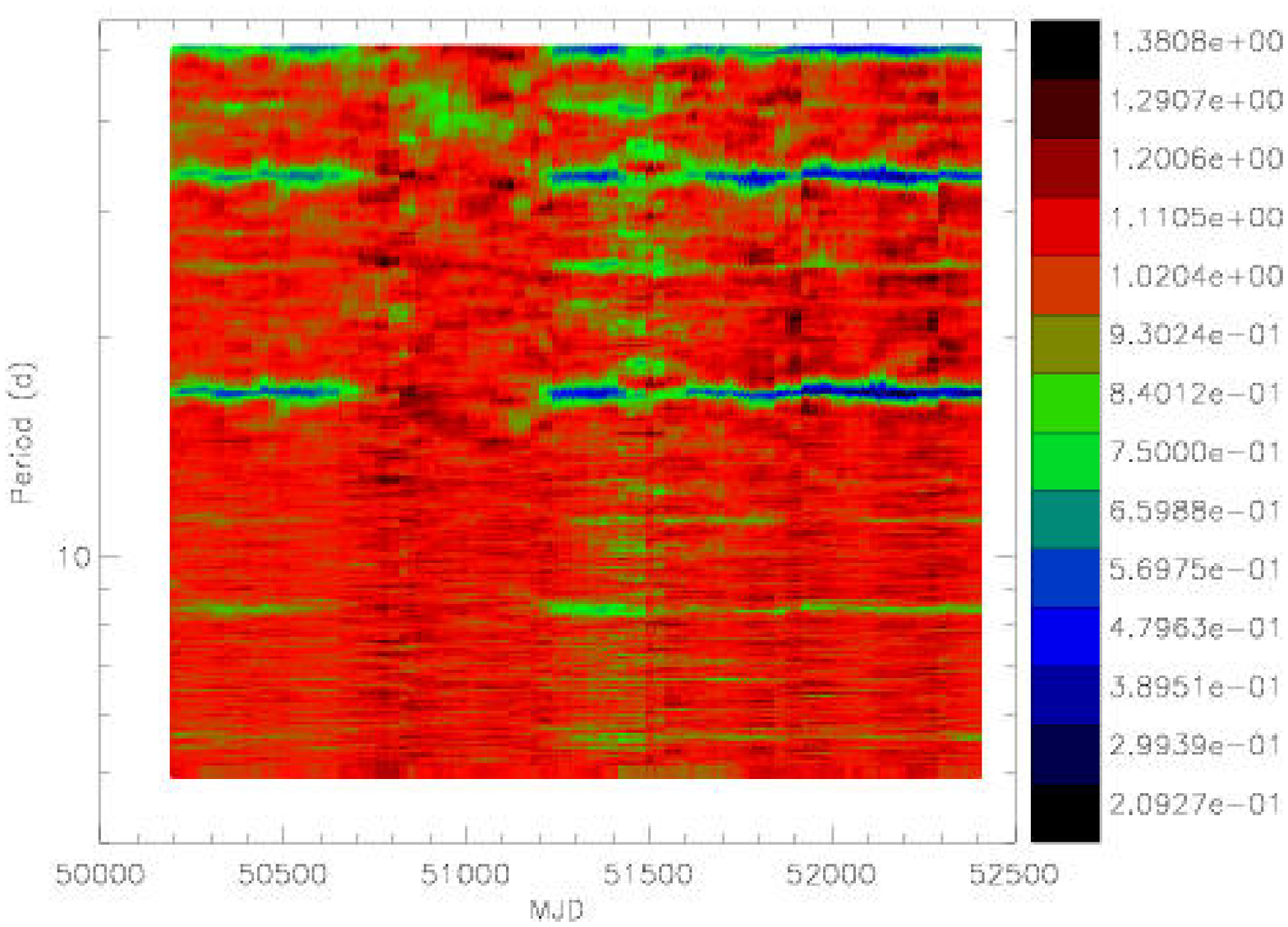}
\caption{Dynamic Periodogram of RXTE ASM data. A sliding window of length $\sim$200 days was stepped through the entire data set in $\sim$20 day increments. {\bf Left} -- Lomb-Scargle dynamic periodogram. {\bf Right} -- PDM dynamic periodogram.\label{fig6}}  
\end{figure}

\begin{figure}
  \begin{center}
    \begin{tabular}{cc}
      	\resizebox{50mm}{!}{\plotone{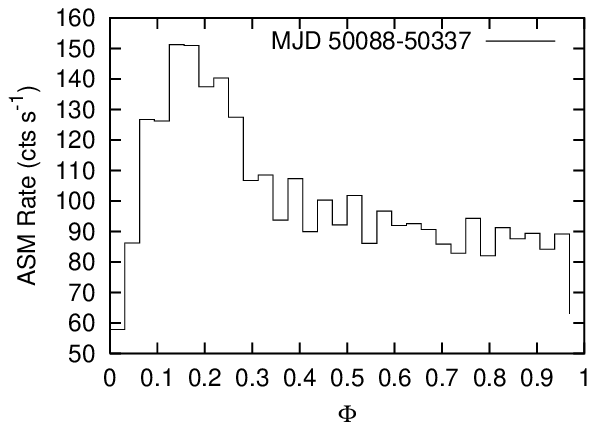}} &
      	\resizebox{50mm}{!}{\plotone{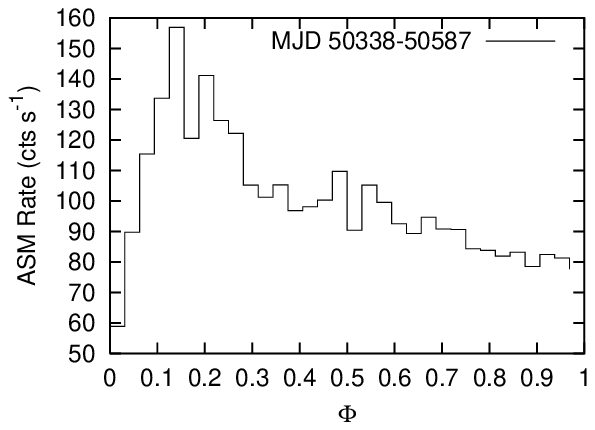}} \\
      	\resizebox{50mm}{!}{\plotone{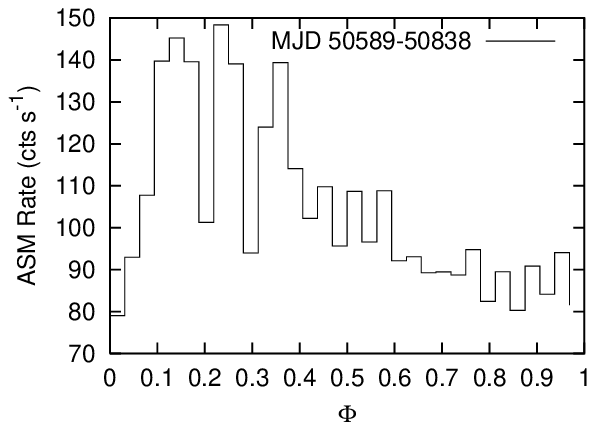}} &
      	\resizebox{50mm}{!}{\plotone{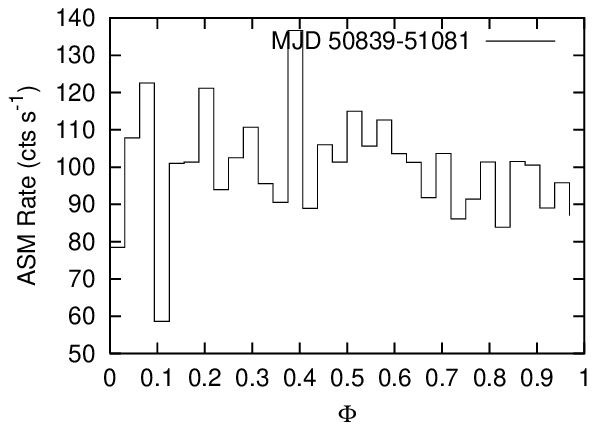}} \\
      	\resizebox{50mm}{!}{\plotone{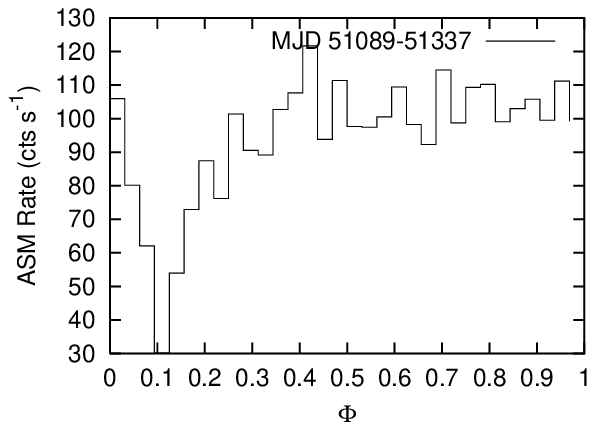}} &
      	\resizebox{50mm}{!}{\plotone{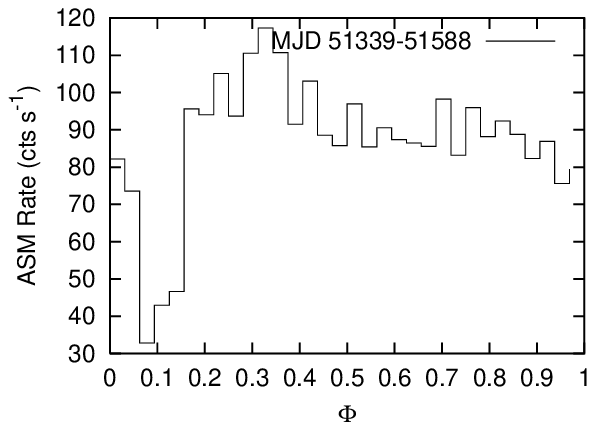}} \\
      	\resizebox{50mm}{!}{\plotone{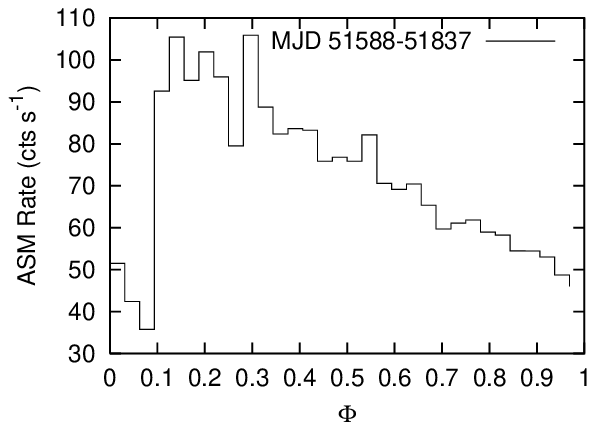}} &
      	\resizebox{50mm}{!}{\plotone{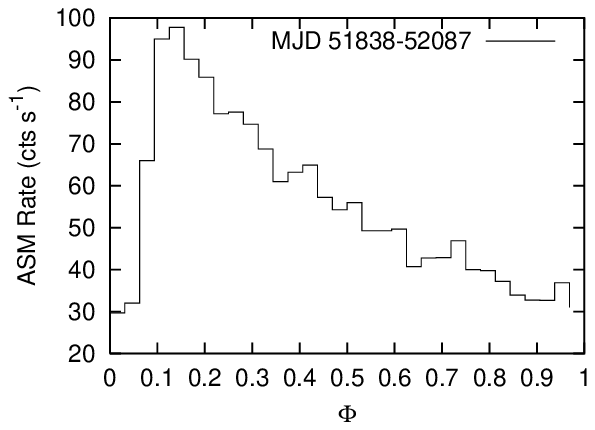}} \\
      	\resizebox{50mm}{!}{\plotone{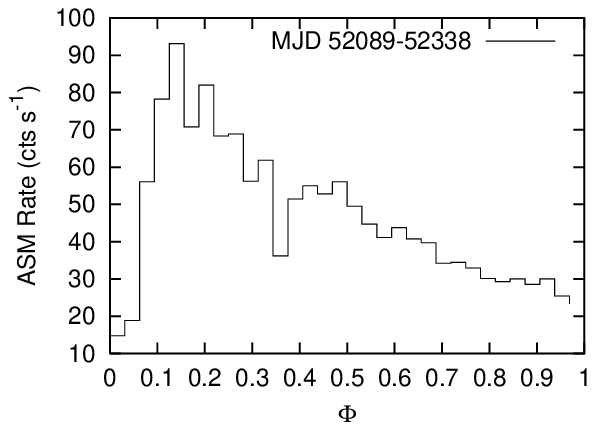}} &
      	\resizebox{50mm}{!}{\plotone{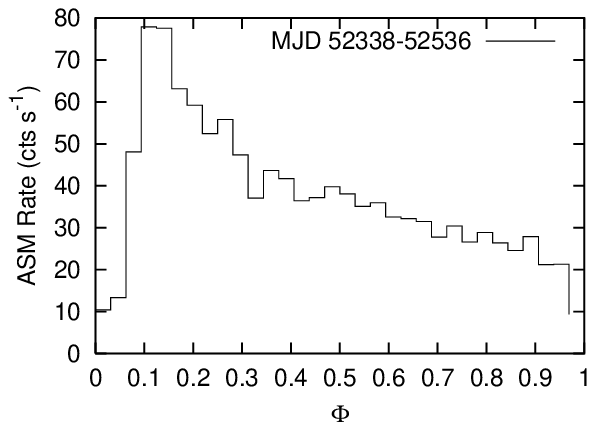}} \\
    \end{tabular}
\caption{RXTE ASM data split into 10 segments ($\sim$250 days long each) and folded at the 16.54 orbital period (Phase 0 
is at MJD 50082). Segments are ordered from top left (MJD 50088--50337) to bottom right (MJD 52338--52536). Segment 4 (MJD 50839--51081) shows no significant modulation at the 16.54 orbital period.\label{fig7}}
\end{center}
\end{figure}

\begin{figure}
\epsscale{1.0}
\plotone{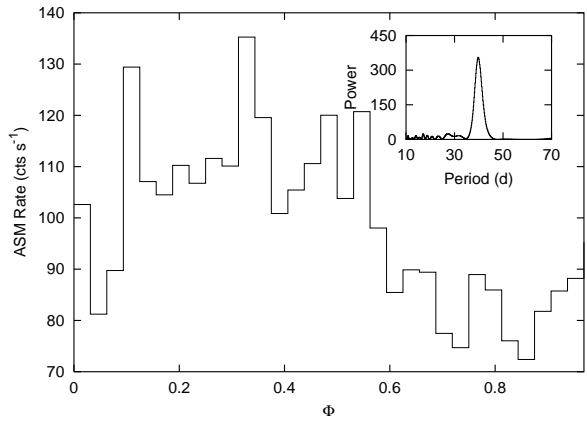}
\caption{{\bf Main Figure:} Segment 4 from Figure \ref{fig7} (MJD 50839--51081) showing a $\sim$40 day modulation. {\bf Inset:} Periodogram showing a peak at 40.0 days (and no significant peak at the orbital period 16.5 d).\label{fig8}}  
\end{figure}

We obtain a characteristic time scale of P/2\.{P} $\sim$1400 years, raising the possibility that this is a much 
younger system than was previously believed. We should note, however, that although we detect a fairly large 
decrease in the orbital period over the course of the more than 30 years of monitoring, we are not able to confirm 
this relatively rapid decrease in period in the RXTE observations alone, which span more than 6.5 years. As we have 
noted (and is clearly visible in the dynamic periodograms in Figure \ref{fig6}), the orbital period of Cir X-1 
was not always detectable during the RXTE observations. In our FFT analysis we split up the data into two distinct
epochs in which the orbital period was believed to be clearly detectable. While two values we obtain show a decrease 
in the orbital period (as would be indicated by the ephemeris), our result, given the errors, is consistent with a 
constant period over the entire RXTE era.

Finally, we have found that there is a period of time of approximately 270 days in which the 16.5 day peak
in the periodograms is suppressed. During this time, in which the presumed orbital period is not 
detectable, a new peak in the periodogram is clearly visible at approximately 40 days. The relationship 
between these two periods is not clear, and the mechanism which might cause this suppression is not 
understood. The dramatic drop in flux level seen for Cir X-1 starting at around MJD 51500, though occurring
about a year later, could be related in some way.

\begin{acknowledgements}
{\noindent \bf Acknowledgements} 
We are grateful to Dr. Shunji Kitamoto for kindly providing the Ginga ASM data used in this paper. We also
thank the referee for useful comments. Results for the ASM light curve provided by the ASM/RXTE teams 
at MIT and at the RXTE SOF and GOF. Much of the analysis in this paper was done using the STARLINK 
package PERIOD, version 4.2. Work at SLAC was supported by Department of Energy contract DE-AC
03-76-SFO0515.  Basic research in X-ray astronomy at the Naval
Research Laboratory is supported by ONR/NRL.  This work was also
supported by the NASA Applied Information Systems Research Program.

\end{acknowledgements}

\def \atel {The Astronomer's Telegram}
\def \apj {ApJ}
\def \apjl {ApJL}
\def \mnras {MNRAS}
\def \iaucirc {IAUCIRC}
\def \em { }
\def \aap {A\&A}
\def \nat {Nature}
\def \araa {Anual Review of Astronomy and Astrophysics}

\clearpage
\twocolumn

\onecolumn

\onecolumn

\onecolumn

\twocolumn

\onecolumn

\twocolumn

\end{document}